\pdfoutput=1
\documentclass[pra,twocolumn,showpacs,floatfix,superscriptaddress]{revtex4-1}
\usepackage{amsfonts,palatino,amsthm,latexsym}
\usepackage[fleqn]{amsmath}
\usepackage{amssymb}
\usepackage{epsfig}
\usepackage{epstopdf}
\usepackage{textcomp}
\usepackage{cancel}
\usepackage{wrapfig}
\usepackage{bbm}
\usepackage[normalem]{ulem}
\usepackage{multirow}
\def\eali\end{align*}
\def\be{\begin{equation}}
\def\ee{\end{equation}}
\def\bea{\begin{eqnarray}}
\def\eea{\end{eqnarray}}
\def\bal{\begin{align}}
\newcommand{\eal}{\end{align}}
\def\ble{\begin{flalign}}
\newcommand{\ele}{\end{flalign}}
\def\ba{\begin{array}}
\def\ea{\end{array}}
\def\bali{\begin{align*}}
\def\nn{\nonumber \\}

\def\bite{\begin{itemize}}

\def\eat{\end{itemize}}
\def\begfig{\begin{figure}[h!]}
\def\endfig{\end{figure}}

\newcommand{\bwe}{\begin{widetext} \begin{eqnarray}}
\newcommand{\ewe}{\end{eqnarray}\end{widetext}}
\newcommand{\ignore}[1]{}

\def\dag{\dagger}

\def\ket#1{\mid #1 {\cal{i}}}

\def\abs#1{\mid #1 \mid}
\def\com#1#2{[#1,#2]}
\def\norm#1{\| #1 \|}

\def\rf#1{Eq.(\ref{#1})}

\def\bd{\bf \begin{definition}: \it}
\def\ed{\end{definition} \rm}
\def\blem{\bf \begin{lemma}: \it}
\def\elem{\end{lemma} \rm}
\def\bthe{\bf \begin{theorem}: \it}
\def\ethe{\end{theorem} \rm}
\def\bcor{\bf \begin{corollary}: \it}
\def\ecor{\end{corollary} \rm}
\def\bpro{\bf \begin{proof}: \rm}
\def\epro{\end{proof} \rm}

\def\totalback{\!\!\!\!\!\!\!\!\!\!\!\!\!\!\!\!}

\def\undback16{& \!\!\!\!\!\!\!\!\!\!\!\!\!\!\!\!}
\def\su2{\ensuremath{\mathfrak{su(2)}}}
\def\SU2{\ensuremath{{SU(2)}}}

\def\+{\ensuremath{\ket{+}}}
\def\-{\ensuremath{\ket{-}}}

\begin{document}
\title{Lieb-Robinson bounds for commutator-bounded operators }

\author{Isabeau Pr\'emont-Schwarz}
\email{ipremont-schwarz@perimeterinstitute.ca}
\affiliation{Perimeter Institute for Theoretical Physics,
31 Caroline St. N,Waterloo ON, N2L 2Y5, Canada}
\affiliation{Department of Physics, University of Waterloo, Waterloo, Ontario N2L 3G1, Canada}
\author{Alioscia Hamma}
\affiliation{Perimeter Institute for Theoretical Physics,
31 Caroline St. N,Waterloo ON, N2L 2Y5, Canada}
\author{Israel Klich}
\affiliation{Department of Physics, University of Virginia,
382 McCormick Rd. Charlottesville, VA 22904, USA}
\author{Fotini Markopoulou-Kalamara}
\affiliation{Perimeter Institute for Theoretical Physics,
31 Caroline St. N, N2L 2Y5, Waterloo ON, Canada}
\affiliation{Department of Physics, University of Waterloo, Waterloo, Ontario N2L 3G1, Canada}

\begin{abstract}
We generalise the Lieb-Robinson theorem to systems whose Hamiltonian is the sum of local operators whose commutators are bounded.
\end{abstract}

\maketitle

The principle of locality is at the heart of the foundations of all modern physics. In quantum field theory, the principle of locality is enforced by an exact light cone. Whenever two (bosonic) observables are spacelike separated, they have to commute, so that neither can have any causal influence on the other. In ordinary quantum mechanics, no explicit request for locality is imposed, and it is, in principle, possible to signal between arbitrarily far apart points in an arbitrarily short time. 
Nevertheless, a simple perturbation analysis shows that such an influence must decay exponentially with the distance between the observables. The seminal work by Lieb and Robinson \cite{lieb} has made this statement rigourous for nonrelativistic spin systems. In essence, it states that any quantum system whose Hilbert space is composed of a tensor product of local, finite-dimensional Hilbert spaces and whose Hamiltonian is the sum of local operators will have an approximately maximum speed of signals. Here, local just means that every operator has as a support the tensor product of few degrees of freedom. The approximation consists of the fact that outside the effective light cone there is an exponentially decaying tail.

Recently, Lieb-Robinson bounds (LRBs) have received renewed interest in both the fields of theoretical condensed matter and quantum information theory \cite{Bravyi:2006zz, eisertetal, cramer, kitaev2003fault, clustering, hastings, Eisert:2006zz, schuch,boso, plenio,sims,lsm,locality,anharmonic}. In particular, the LRB has been used to prove that a nonvanishing spectral gap implies an exponential clustering in the ground state \cite{hastings, clustering, schuch}. Further developments can be found in \cite{sims}, where the LRB is used also to argue about the existence of dynamics. The LRB has also been instrumental in the recent extension of the Lieb-Schultz-Mattis theorem to higher dimensions \cite{hastingsa,lsm}. In \cite{Eisert:2006zz,hastingsb}, it has been shown how the Lieb-Robinson bounds can be exploited to find general scaling laws for entanglement. In \cite{Bravyi:2006zz}, these techniques have been exploited to characterise the creation of topological order. The locality of dynamics has important consequences on the simulability of quantum spin systems. In \cite{osborne1,osborne2} it has been shown that one-dimensional gapped spin systems can be efficiently simulated. A review of some of the most relevant aspects of the locality of dynamics for quantum spins systems can be found in \cite{locality}. Other developments of significant interest include \cite{boso,plenio} which show that it is possible to entangle macroscopically separated nanoelectromechanical oscillators of the oscillator chain and that the resulting entanglement is robust to decoherence. Such a system is of great interest for its possible application as a quantum channel and as a tool to investigate the boundary between the classical and quantum worlds.

The LRBs have found a more exotic use in the field of emergent gravity, where one wants to study locality, geometry and Lorentz symmetry as emergent phenomena \cite{Konopka:2008hp,Konopka:2008ds}. An example of the usefulness of the Lieb-Robinson bounds can be found in \cite{LRB1}, where it was shown that in spin systems with emergent electromagnetism \cite{wenlight}, the speed of light is also the maximum speed of signals, without imposing from the beginning any Lorentz invariance. This raises the issue whether even Lorentz invariance could be emergent.

One problem with the Lieb-Robinson bounds is that it is 
difficult to obtain bounds for unbounded Hamiltonians.  In the usual Lieb-Robinson settings, the Hamiltonian must be a sum of local \emph{bounded} operators. If the unbounded terms in the Hamiltonian are completely local, that is, if they are on-site terms, it is possible to prove a Lieb-Robinson theorem using the usual technique \cite{anharmonic}. In the specific case of coupled harmonic oscillators on a graph with local interactions, it was proven in \cite{eisertetal} that the Lieb-Robinson bound is valid for canonical and Weyl operators and a proof for a generalisation to general operators is outlined. Algebraic suppression (instead of the usual exponential suppression of the Lieb-Robinson Bound) is shown to result from nonlocal algebraic interactions. As an interesting corollary, \cite{eisertetal} shows how the approximate locality implied by the Lieb-Robinson bound becomes exact in the continuous limit for the Klein-Gordon field.

In this paper, we will show how one can find a bound to the maximum speed of interactions in the case of a class of unbounded spin Hamiltonians. It is not true that for any unbounded Hamiltonian, a Lieb-Robinson bound exists \cite{supersonic}. Here, we want to show that one can derive a Lieb-Robinson bound if the Hamiltonian is the sum of local operators, \emph{whose commutators are bounded}. Therefore, there is no necessity for even the nonlocal terms to be bounded, as long as their commutators are. More specifically, we show that for quantum systems whose Hilbert space is the tensor product of local Hilbert spaces associated with vertices and edges of a graph, if the Hamiltonian is the sum of local operators $\Phi_i$, each with a support on a region of the graph with a diameter less than a fixed number $R$,if each of these local operators $\Phi_i$ is noncommuting with less than $\nu$ other local operator terms of the Hamiltonian $\Phi_j$, and if for any two of these operators we have $\norm{\com{\Phi_i}{\Phi_j}}<K$ and for any three operators $\norm{\com{\Phi_i}{\com{\Phi_j}{\Phi_k}}}<Q$ for two positive numbers $K$ and $Q$, then we have that for any two local operators $\Phi_i$ and $\Phi_i$ which are terms in the Hamiltonian and whose support is separated by a graph distance $d$, that
\bea
\norm{\com{\Phi_i(t)}{\Phi_j(0)}}  \leq \tilde{\tilde{M}} \exp{\Big[\lambda\left(v_{LR} t-  d\right)\Big]} ,\label{almostf}
\eea
where $\tilde{\tilde{M}}$ is a constant and $v_{LR}$, the limit on the speed of propagation of information, depends only on local operators of the Hamiltonian as it is they who affect the propagation. This bound can be generalised to any local observables $O_P$ and $O_Q$ with supports $P$ and $Q$ respectively, that satisfy the following local observable operator conditions:$(i)$ The graph distance $d$ separating $P$ and $Q$ is greater than $R$. $(ii)$ The number of terms $\Phi_i$ of the Hamiltonian whose support has nonempty intersection with $P$ is $n_P<\infty$.  $(iii)$ There exists $F_P$ and $F_Q$ such that for all terms $\Phi_i$ and $\Phi_j$ of the Hamiltonian the inequalities $\norm{\com{O_P}{\Phi_i}}<F_P K$, $\norm{\com{O_Q}{\Phi_i}}< F_Q K$ and $\norm{\com{O_Q}{\com{\Phi_i}{\Phi_j}}}<F_Q Q$ are satisfied. The generalised bound is then
\bea
\norm{\com{O_P(t)}{O_Q(0)}} && \nn\leq  \tilde{\tilde{M}} F_P F_Q &&  n_P(n_P+1)  \exp{\Big[\lambda\left(v_{LR} t-  d\right)\Big]} . \label{finalb}
\eea

To motivate our discussion, let us start by the most trivial example: Consider the case of a
Hamiltonian $H=\sum h_i$ which is composed of a sum of local terms $h_i$ which are commuting, such as the quantum Ising model without the transverse field. 
In such a case, there is simply no
propagation of signals: Indeed, for any local operator $O_A$ we have $O_A(t)=e^{itH}O_Ae^{-iHt}=
e^{itH_A}O_Ae^{-iH_At}$, where $H_A=\sum_{i:[h_i,O_A]\neq 0} h_i$, since there is a finite number of $h_i$ in $H_A$, and they are of finite range; $O_A(t)$ is also strictly local for arbitrary long times $t$, irrespective of the norm of the $h_i$ operators.
This suggests that it is desirable to find Lieb-Robinson bounds in terms of the norm of the commutators rather than the norm of the local terms $h_i$.

Let us outline a simple example. Consider a system of parallel quantum wires. We place fermions on the wires, and these are usually described by one-dimensional Luttinger liquids, and have approximately a linear dispersion relation. We place a density-density interaction between the wires. Labeling the wires by the index $j$, the system can be described by the following Hamiltonian:
\begin{equation}
H_{wires} =  \sum_j \left(-i\frac{\partial}{\partial x_j} + V(x_j - x_{j+1})\right),
\end{equation}
which is commutator bounded in the sense of this paper as long as both $| \partial_{x_j} V|$ and $| \partial_{x_i}\partial_{x_j} V|$ are bounded. Another example involves a generalised Dicke model, describing an array of spins interacting with a boson field via
\begin{eqnarray}
H=\sum h_n\,\,\,\,;\,\,\,\, h_n=\sigma^z_n(b_n^{\dag}+b_n+ib_{n+1}^{\dag}-ib_{n+1})
\end{eqnarray}
where $b_n$ are boson creation operators and $\sigma_n^z$ is the $n$th spin. It is easy to check that in this case the commutator $[h_n,h_{n+1}]=-2 i \sigma^z_n\sigma^z_{n+1}$ is bounded. [In fact, this particular Hamiltonian can also be written as a sum of commuting terms $\tilde{h}_n=b_n(\sigma^z_n-i\sigma^z_{n-1})+h.c.$].

We consider Hamiltonians that are the sum of two different types of operators $\Phi_0$ and $\Phi_1$ :
\begin{align}
H \equiv \sum_{i\in S_0} h_0 \Phi_0^i + \sum_{j\in S_1} h_1 \Phi_1^j \label{deuxint}.
\end{align}
Here, $S_0,S_1$ are two sets of labels, $h_0$ and $h_1$ are two coupling constants, and $[\Phi_0^i,\Phi_0^j]=[\Phi_1^i,\Phi_1^j]=0$ for every $i,j$. As an example, consider the Ising model. Then $\Phi_0^i = \sigma^x_i \sigma^x_{i+1}$ and $\Phi_1^i = \sigma^z_i$. We call the subgraph which is the support of the operator $\Phi_q^m,\ \Gamma(q,m)$ and for $(a,b)\in\{0,1\}^2$, we define
\begin{align}
K_{a \ b}^{i \ j}(t) \equiv \com{\Phi_a^{i}(t)}{\Phi_b^j} . \label{f}
\end{align}

We consider what we will refer to as commutator-bounded $R$-local quantum systems.  For such systems, the commutators and the commutators of commutators of operators of the Hamiltonian are uniformly bounded while the operators themselves may be unbounded and the operators of the Hamiltonian have support on subgraphs of size less than $R$, for $R$ an arbitrary natural number.  Explicitly, the diameter of all $\Gamma(q,m)$ is less than $R$ and for any three operators $\Phi_a^i$, $\Phi_b^j$, and $\Phi_c^k$ appearing in the Hamiltonian with the coupling constants $h_a$, $h_b$, and $h_c$, we have that $ h_a h_b\norm{\com{\Phi_a^i}{\Phi_b^j}}< K\ $ and  $h_a h_b h_c\norm{\com{\com{\Phi_a^i}{\Phi_b^j}}{\Phi_c^k}}< Q$, where $K$ and $Q$ are positive real numbers. Note that a bounded system, which is uniformly bounded by $\tilde{K}$, must satisfy $K\leq 2\tilde{K}^2$ as well as $Q\leq 4\tilde{K}^3$, and thus boundedness implies commutatorboundedness.

By taking the derivative of \rf{f} with respect to $t$, we obtain $({K_{a \ b}^{i_1 \ j}}(t))^\prime =  \com{\com{-i H(t)}{\Phi_a^{i}(t)}}{\Phi_b^j}$, after keeping only the terms in $H(t)$ which do not commute with $\Phi_a^{i}(t)$,  and after some algebra, we get [here and in the following by $a+1$ we mean  $a+1\ mod(2)$]
\begin{eqnarray}
\nonumber
 ({K_{a \ b}^{i_1 \ j}}(t))^\prime =  \com{{K_{a \ b}^{i_1 \ j}}(t)}{\Big(-i h_{a+1} \sum_{i_2\in Z_{i_1}} \Phi_{a+1}^{i_2}(t)\Big)} \\ +  (-i h_{a+1}) \sum_{i_2\in Z_{i_1}}\com{\Phi_a^{i_1}(t)}{\com{\Phi_{a+1}^{i_2}(t)}{\Phi_b^j}},
\label{fpp}
\end{eqnarray}
where, if $i\in S_0$, then $Z_i$ is the finite subset of $S_1$ such that $j\in Z_i \Leftrightarrow \Gamma(0,i)\bigcap \Gamma(1,j)\neq \emptyset$ and vice versa for $i\in S_1$.

Taking the second derivative, and using the fact that $\com{\Phi_{a}^{i}(t)}{\Phi_{a}^{j}(t)} = 0$, we obtain, after some algebraic manipulation,
\begin{widetext}
\begin{eqnarray}
\nonumber
 {K_{a \ b}^{i_1 \ j}}''(t)   = &&  -i \com{{K_{a \ b}^{i_1 \ j}}'(t)}{\sum_{j_2\in Z_{i_1}} h_{a+1}\Phi_{a+1}^{j_2}(t) +  \sum_{i_3\in Z_{i_2}} h_{a}\Phi_{a}^{i_3}(t)}  -  h_{a+1}^2 \sum_{j_2\in Z_{i_1}}\sum_{i_2\in Z_{i_1}} \com{\com{\Phi_a^{i_1}(t)}{\Phi_{a+1}^{i_2}(t)}}{\com{\Phi_{a+1}^{j_2}(t)}{\Phi_b^j}} \\  &&- h_{a} h_{a+1}\sum_{i_2\in Z_{i_1}}\sum_{i_3\in Z_{i_2}} \com{\com{\Phi_a^{i_1}(t)}{\Phi_{a+1}^{i_2}(t)}}{\com{\Phi_{a}^{i_3}(t)}{\Phi_b^j}}. \label{fqtt}
\end{eqnarray}
\end{widetext}
Defining the following unitary operator $U_{2+3}(t)\equiv e^{-it \left(\sum_{j_2\in Z_{i_1}} h_{a+1}\Phi_{a+1}^{j_2}(t) +  \sum_{i_3\in Z_{i_2}} h_{a}\Phi_{a}^{i_3}(t)\right)}$ and its associated unitary evolution $T_{2+3}(t) O \equiv U_{2+3}(t)^\dag O U_{2+3}(t)$, integrating \rf{fqtt},  and taking the norm, we obtain, after some manipulations,
\begin{widetext}
\begin{eqnarray}
\nonumber
\norm{{K_{a \ b}^{i_1 \ j}}'(t)}   \leq && h_{a+1} \sum_{i_2\in Z_{i_1}} \norm{\com{\com{\Phi_a^{i_1}}{\Phi_{a+1}^{i_2}}}{\Phi_b^j}} \\
&& +\int_0^t ds \left( 2 h_{a+1}^2K \sum_{j_2\in Z_{i_1}}\sum_{i_2\in Z_{i_1}} \norm{\com{\Phi_{a+1}^{j_2}(s)}{\Phi_b^j}} \right.  +  \left. 2 h_{a} h_{a+1} K \sum_{i_2\in Z_{i_1}}\sum_{i_3\in Z_{i_2}}\norm{\com{\Phi_{a}^{i_3}(s)}{\Phi_b^j}}\right), \label{leqy}
\end{eqnarray}
where, we used the fact that $\norm{\com{\Phi_a^{i}(t)}{\Phi_{a+1}^{j}(t)}}\leq K$.  By integrating \rf{leqy}, we get
\begin{align}
\nonumber
\norm{K_{a b}^{i_1 j}(t)} & \leq   \norm{\com{\Phi_a^{i_1}}{\Phi_b^j}} + h_{a+1} \sum_{i_2\in Z_{i_1}}\norm{\com{\com{\Phi_a^{i_1}}{\Phi_{a+1}^{i_2}}}{\Phi_b^j}} t \\ & +  \int_0^t ds\int_0^s dl \left( 2 h_{a+1}^2 K \sum_{i_2\in Z_{i_1}}\sum_{j_2\in Z_{i_1}} \norm{K_{a+1 b}^{j_2 j}(l)}
  +  2 h_{a} h_{a+1} K \sum_{i_2\in Z_{i_1}}\sum_{i_3\in Z_{i_2}} \norm{K_{a b}^{i_3 j}(l)}\right). \label{KUN}
\end{align}
Since the commutators are bounded, we have $\norm{\com{\Phi_a^{i_1}}{\Phi_b^j}} \leq K$ and $\norm{\com{\com{\Phi_a^{i_1}}{\Phi_{a+1}^{i_2}}}{\Phi_b^j}} \leq Q$ for some $K,Q>0$.  Noting that $\norm{\com{\Phi_a^{i_1}}{\Phi_b^j}}= 0$ if $\Gamma(a,i_1)$ and $\Gamma(b,j)$ do not overlap and $\norm{\com{\com{\Phi_a^{i_1}}{\Phi_{a+1}^{i_2}}}{\Phi_b^j}} =0$ if $\Gamma(b,j)$ does not overlap with either $\Gamma(a,i_1)$ or $\Gamma(a+1,i_2)$, we see that \rf{KUN} implies
\begin{align}
 \norm{K_{a b}^{i_1 j}(t)} & \leq  K \delta_{i_1}^{j} + \sum_{i_2\in Z_{i_1}}h_{a+1} Q \delta_{i_1\cup i_2}^{j} t  +  \int_0^t ds\int_0^s dl \left( 2 h_{a+1}^2  \sum_{j_2\in Z_{i_1}}\sum_{i_2\in Z_{i_1}}\norm{K_{a+1 b}^{j_2 j}(l) } +  2 h_{a} h_{a+1} \sum_{i_2\in Z_{i_1}}\sum_{i_3\in Z_{i_2}}\norm{ K_{a b}^{i_3 j}(l)}\right), \label{nino}
\end{align}
where we have used the following symbol:
\begin{align}
\delta_{i}^k := \left\{ \begin{array}{cc} 1 & \mbox{if } \Gamma(a_i,i)\cap \Gamma(a_k,k)\neq \emptyset , \\ 0 & \mbox{otherwise}. \end{array}\right. \label{eqn_delta}
\end{align}
Solving for $ \norm{K_{a b}^{i_1 j}(t)}$, we find
\begin{align}
 \norm{K_{a b}^{i_1 j}(t)}  & \leq   K \delta_{i_1}^{ j} + \sum_{i_2\in Z_{i_1}} Q h_{a+1} \delta_{i_1\cup i_2}^{ j} t
 +  \sum_{j_2\in Z_{i_1}}\sum_{i_2\in Z_{i_1}} 2 h_{a+1}^2 K \delta_{j_2}^{ j} \frac{t^2}{2!}
 +  \sum_{i_2\in Z_{i_1}}\sum_{i_3\in Z_{i_2}} 2 h_{a} h_{a+1} K \delta_{i_3}^{ j} \frac{t^2}{2!}\nn
& +   \sum_{j_2\in Z_{i_1}}\sum_{i_2\in Z_{i_1}}\sum_{i_3\in Z_{j_2}} 2 h_{a+1}^2 h_a K Q \delta_{j_2\cup i_3}^{ j} \frac{t^3}{3!} +
 \sum_{i_2\in Z_{i_1}}\sum_{i_3\in Z_{i_2}}\sum_{i_4\in Z_{i_3}} 2 h_{a+1} h_a^2 K Q \delta_{i_3\cup i_4}^{ j} \frac{t^3}{3!} \nn
& + \int_0^t dv\int_0^v du\int_0^u ds\int_0^s dl \Bigg(
 (2 h_{a+1}h_a)^2 K^2 \sum_{i_2\in Z_{i_1}}\sum_{j_2\in Z_{i_1}}\sum_{i_3\in Z_{j_2}}\sum_{j_3\in Z_{j_2}} \norm{K_{a b}^{j_3 j}(l)}\nn
&  + \ (2 h_{a+1}^2)(2 h_{a+1}h_a) K^2\sum_{i_2\in Z_{i_1}}\sum_{j_2\in Z_{i_1}}\sum_{i_3\in Z_{j_2}}\sum_{i_4\in Z_{i_4}} \norm{K_{a+1 b}^{j_4 j}(l)}\nn
& +  (2 h_{a+1}h_a)^2 K^2 \sum_{i_2\in Z_{i_1}}\sum_{i_2\in Z_{i_1}}\sum_{i_3\in Z_{i_2}}\sum_{i_4\in Z_{i_3}} \norm{K_{a+1 b}^{i_4 j}(l)}    +  \left. (2 h_{a}^2)(2 h_{a+1}h_a) K^2 \sum_{i_2\in Z_{i_1}}\sum_{i_3\in Z_{i_2}}\sum_{i_3\in Z_{1_2}}\sum_{j_3\in Z_{i_2}} \norm{K_{a b}^{j_3 j}(l)}\right).\label{K3}
\end{align}
\end{widetext}
Iterating this procedure we obtain by induction
\begin{align}
\norm{K_{a b}^{i_1 j}(t)} \leq M \sum_{n=0}^\infty \sqrt{2 h_0 h_1 K}^n \frac{\abs{t}^n}{n!} c_n \label{K40},
\end{align}
where $M = \sqrt{2} \max\{\frac{h_0}{h_1},\frac{h_1}{h_0}\}\times\max\{\frac{1}{K},1\}\times\max\{\frac{\sqrt{K}}{Q},1\}$ and where $c_n$ is a combinatorial factor counting the number of linking operator chains of $n$ operators between $\Gamma(a,i_1)\ $and $\Gamma(b,j)$. What we call an operator chain is heuristically a sequence of intersecting operators linking the initial and final operators.
 The process of constructing the sequence of operators forming the chain is as follows: The $2j^{th}$ operator in the chain has to be noncommuting with the $(2j-1)^{th}$ one. This imposes that the two consecutive operators of a chain have to (1) be of a different interaction type and (2) have overlapping support. For the odd-numbered operators, there is an extra choice: The $(2k+1)^{th}$ operator can be an operator that does not commute with the $2k^{th}$ operator (as for the even case) \emph{or} \rm an operator that does not commute with the $(2k-1)^{th}$ operator. That is, even operators in the sequence must be noncommuting with the previous operator in the sequence and odd operators in the sequence must be noncommuting with either of the two previous operators in the sequence.  From the recursive \rf{nino}, we see that if we start with an operator $i_1$ of type $a$, the next operator in the chain must be an operator $i_2\in Z_{i_1}$ of the other type ($a+1$). The fact that it is in $Z_{i_1}$ means that its support overlaps with $i_1$'s, which is similar to what was found in the bounded case. However if we look at the operator that comes after $i_2$, we see that we have two distinct possibilities. The first [second double sum under the integrals of \rf{nino}] is that it can be $i_3\in Z_{i_2}$ an operator of type $a$ (different from $a+1$) whose support overlaps with $i_2$'s; if this were the only possibility, we would have exactly the same situation as we had for bounded systems, but the first double sum under the integrals of \rf{nino} adds another possibility. That second possibility (first double sum) is choosing an operator $j_2\in Z_{i_1}$ after the operator $i_2$, $j_2$ is, like $i_2$, an operator of type $a+1$ which (by virtue of being in $Z_{i_1}$) has a support overlapping $i_1$'s. To find the next operator after that, we reiterate \rf{nino} and thus, like the first operator after $i_1$, we need to choose an operator which is of a different type than the last one (be it $j_2$ or $i_3$) and has overlapping support with the last one; thus, at this point, we cannot ``change our mind".  We can thus see the process of building the chain as, for every two choices, we must choose an operator that links with the previous one, but every other choice, we can also choose an operator that links to the penultimate one instead.

Because every two choices in building up the chain we must choose an operator of a different type than the previous one, in the end, the chain contains the same number of operators of type $0$ as of type $1$ (plus or minus one). This means that there will be the same number of factors of $h_1$ as of $h_0$ in every term; hence, we can pull them out of the sums over chains and simply write an overall factor of $\sqrt{h_0 h_1}$ in front while passing from \rf{K3} to \rf{K40}.

Furthermore, we can always find a bound of the following type for $c_n$:
\bea
c_n \leq \tilde{M} \gamma^n e^{\lambda(\frac{n}{\xi} - d)} \label{cng1},
\eea
where $\lambda$ is an arbitrary positive real number.
This is because the $\Gamma(a,i)$'s have a diameter of $R$ or less. Hence, if the distance $d$ between the initial and final points is greater than $R^n$, then there are no possible linking operator chains of $n$ local operators between the initial and final points. Furthermore, since at every odd step along the chain there is a choice of at most $\nu$ local operators to choose from for the next operator in the chain and at every even step there is at maximum $2\nu$ operators to choose from,
there is, at most, $(\sqrt{2}\nu)^n$ possible local operator chains of $n$ operators starting from any given position.  Thus, we certainly have that
\bea
c_n \leq \sqrt{2}^n \nu^n e^{\lambda(R {n} - d)} \label{cngex},
\eea
where $\lambda$ is arbitrary.
Using \rf{cng1} with \rf{K40}, we obtain the LRB of \rf{almostf},
\bea
\totalback \norm{\com{\Phi_a^{i}(t)}{\Phi_b^j(0)}}  \leq \tilde{\tilde{M}} \exp{\lambda\left(2\sqrt{h_0 h_1 K}\frac{\gamma}{\lambda} e^{\frac{\lambda}{\xi}}t-  d\right)} ,\label{LRBUB}
\eea
where $\tilde{\tilde{M}} = \tilde{M}M$. To obtain the generalisation to local operators $O_P$ and $O_Q$ satisfying the local observable operator conditions enounced in the introduction, we introduce $\tilde{K}_{a}^{i_1}(t)\equiv \com{\Phi_a^{i}(t)}{O_Q(0)}$. Using exactly the same procedure used to obtain \rf{leqy}, we get
\bea
&& \norm{\com{O_P(t)}{O_Q(0)}'(t)}   \leq \nn
&& \int_0^t ds \left( 2 h_{b}h_{a} K F_P \sum_{j_2\in Z_{P}}\sum_{i_2\in Z_{P}} \norm{\tilde{K}_{b}^{j_2}(s)} \right. \nn && +  \left. 2 h_{a} h_{a+1} K F_P \sum_{i_2\in Z_{P}}\sum_{i_3\in Z_{i_2}}\norm{\tilde{K}_{a}^{i_3}(s)}\right)\nn && \leq 2 \max\{h_0^2, h_1^2\} n_P(n_P+1)\int_0^t ds \norm{\tilde{K}_{a}^{k}(s)} , \label{leqyfier}
\eea
 where $Z_P$ is the set of of labels of the terms of the Hamiltonian which do not commute with $O_P$ ($\norm{Z_P}=n_P$), where $k$ is such that $\int_0^t ds \norm{\tilde{K}_{a}^{k}(s)} = \max_{i\in Z_P} \int_0^t ds \norm{\tilde{K}_{a}^{i}(s)}$ and where, unlike in \rf{leqy}, the terms containing no integrals do not appear here because of the condition that $d>R$. $\norm{\tilde{K}_{a}^{k}(s)}$ can then be treated in exactly the same way as $ \norm{K_{a b}^{i_1 j}(s)}$ was, with the only exception that while bounding the final commutators [i.e., when we place the $\delta$ of \rf{eqn_delta}], we will need an extra factor of $F_Q$. Thus, we obtain \rf{finalb}:
\bea
\totalback \norm{\com{O_P(t)}{O_Q(0)}}  \leq && \nn F_P F_Q n_P(n_P && +1)\tilde{\tilde{M}} \exp{\lambda\left(2 \sqrt{h_0 h_1 K}\frac{\gamma}{\lambda} e^{\frac{\lambda}{\xi}}t-  d\right)} .\label{LRBUBG}
\eea
Optimising for $\lambda$, we have that the Lieb-Robinson speed is thus
\bea
v_{L R} = 2\frac{\gamma}{\xi} e\sqrt{h_0 h_1 K} .\label{vlrunb}
\eea
We can compare \rf{LRBUBG} with the bound obtained for Hamiltonians composed of bounded local operators and for bounded local observables $O_P$ and $O_Q$, which is \cite{LRB1}
\bea
\totalback \norm{\com{O_P(t)}{O_Q(0)}}  \leq && \nn \norm{O_P}\norm{O_Q}n_P && \tilde{\tilde{M}} \exp{\lambda\left(2 \sqrt{h_0 h_1 }\frac{\gamma}{\lambda} e^{\frac{\lambda}{\xi}}t-  d\right)} .\label{LRBUBounded}
\eea

To summarize, in this paper, we have shown that a Lieb-Robinson bound exists for those Hamiltonians that are the sum of local operators whose commutator is bounded. This allows for treating a class of systems with unbounded operators.

\emph{Acknowledgments.---} Research at Perimeter Institute for Theoretical
Physics is supported in part by the Government of Canada through NSERC and
by the Province of Ontario through MRI.


\begin{thebibliography}{99}
\vspace{-2cm}
\linespread{0.9}
\bibitem{lieb} E.~H.~Lieb, and D.~W.~Robinson, Comm. Math. Phys. \textbf{28},
251 (1972).
\bibitem{Bravyi:2006zz}
  S.~Bravyi, M.B.Hastings, and F. Verstraete,\prl  {\bf 97}, 050401 (2006).
\bibitem{eisertetal} M. Cramer, A. Serafini, J. Eisert, \emph{Quantum Information and Many Body Quantum Systems},
M. Ericsson and S. Montangero eds, Edizioni della Normale, Pisa,pp. 51-72
2008. [arXiv:0803.0890v2].
\bibitem{cramer}
  M.,~Cramer and J.,~Eisert,  New J. Phys. 8, 71 (2006)
 \bibitem{kitaev2003fault}
A.~Kitaev, Ann. Phys. (NY), 303 (2003) [arXiv:quant-ph/9707021]

  \bibitem{clustering} B.~Nachtergaele et al., Comm. Math. Phys. {\bf 265}, 119 (2006).
\bibitem{Eisert:2006zz}
  J.~Eisert et al.,
  Phys.\ Rev.\ Lett.\  {\bf 97}, 150404  (2006) .
\bibitem{schuch}
N.~Schuch et al.,
Comm. Math. Phys. {\bf 267}, 65 (2006).
\bibitem{sims} B.~Nachtergaele, Y.~Ogata, and R.~Sims, J. Stat. Phys.
\textbf{124}, 1 (2006)
[arXiv:math-ph/0603064v1].
\bibitem{anharmonic} B.~Nachtergaele, H.~Raz, B.~Schlein, and R.~Sims, Comm. Math. Phys. {\bf 286}, 1073-1098 (2009)
\bibitem{hastingsa} M. B.~Hastings, \prb {\bf 69}, 104431 (2004)
\bibitem{hastingsb} M. B.~Hastings, \prl {\bf 93}, 140402 (2004) [arXiv:cond-mat/0406348].
\bibitem{hastings} M.B.~Hastings et al., Comm. Math. Phys. \textbf{265},
781 (2006).
\bibitem{lsm} B.~Nachtergaele,  and R.~Sims, Comm. Math. Phys. {\bf 276}, 437-472 (2007).
\bibitem{locality} B.~Nachtergaele,  and R.~Sims, [arXiv:0712.3318].
\bibitem{boso}
J. Eisert et al., Phys. Rev. Lett. \textbf{93}, 190402 (2004).
\bibitem{plenio} M.B. Plenio, et al., New J. Phys. \textbf{6}, 36 (2004).
\bibitem{supersonic} J. Eisert, D. Gross, 	arXiv:0808.3581v3
\bibitem{wenlight} X.-G.~Wen, \prd {\bf 68}, 065003 (2003).
\bibitem{osborne1} T.J.~Osborne, Phys. Rev. Lett. {\bf 97}, 157202 (2006)
\bibitem{osborne2} T.J.~Osborne, Phys. Rev. A {\bf 75}, 032321 (2007)
\bibitem{Konopka:2008hp}
  T.~Konopka, F.~Markopoulou and S.~Severini,
  Phys.\ Rev.\  D {\bf 77}, 104029 (2008).

\bibitem{Konopka:2008ds}
  T.~Konopka,
  Phys.\ Rev.\  D {\bf 78}, 044032  (2008).

\bibitem{LRB1}
  A.~Hamma, F.~Markopoulou, I.~Pr\'emont-Schwarz and S.~Severini,
  Phys.\ Rev.\ Lett.\  {\bf 102}, 017204  (2009).



\end{thebibliography}
\end{document}